# DeepNIS: Deep Neural Network for Nonlinear Electromagnetic Inverse Scattering

Lianlin Li, *Senior Member*; Long Gang Wang; Fernando L. Teixeira, *Fellow*; Che Liu; Arye Nehorai, *Life Fellow*; and Tie Jun Cui, *Fellow*

*Abstract*—Nonlinear electromagnetic (EM) inverse scattering is a quantitative and super-resolution imaging technique, in which more realistic interactions between the internal structure of scene and EM wavefield are taken into account in the imaging procedure, in contrast to conventional tomography. However, it poses important challenges arising from its intrinsic strong nonlinearity, ill-posedness, and expensive computation costs. To tackle these difficulties, we, for the first time to our best knowledge, exploit a connection between the deep neural network (DNN) architecture and the iterative method of nonlinear EM inverse scattering. This enables the development of a novel DNN-based methodology for nonlinear EM inverse problems (termed here DeepNIS). The proposed DeepNIS consists of a cascade of multi-layer complex-valued residual convolutional neural network (CNN) modules. We numerically and experimentally demonstrate that the DeepNIS outperforms remarkably conventional nonlinear inverse scattering methods in terms of both the image quality and computational time. We show that DeepNIS can learn a general model approximating the underlying EM inverse scattering system. It is expected that the DeepNIS will serve as powerful tool in treating highly nonlinear EM inverse scattering problems over different frequency bands, involving large-scale and high-contrast objects, which are extremely hard and impractical to solve using conventional inverse scattering methods.

## I. INTRODUCTION

A wide range of scientific, engineering, military, and medical applications benefit from nonlinear electromagnetic (EM) inverse scattering as an accurate, non-destructive imaging reconstruction tool [1-6].

L. Li and L. Wang are with the School of Electronics Engineering and Computer Sciences, Peking University, Beijing, 100871, China
lianlin.li@pku.edu.cn
F. L. Teixeira is with the ElectroScience Laboratory, The Ohio State University, Columbus OH, 43212, USA; teixeira.5@osu.edu
C. Liu and T. J. Cui are with the State Key Laboratory of Millimeter Waves, Southeast University, Nanjing 210096, China;
A. Neorail is with the Department of Electrical & Systems Engineering, Washington University in St. Louis, Washington, 63130, USA.

As the nonlinear EM inverse scattering is capable of accounting for multiple scattering of EM wavefields inside the scene [3-7], one can "see" the internal structure of scene in a quantitative way that is superior to the conventional tomography methods [8-9, 35-37]. In the past decades, a plethora of EM inverse scattering algorithms have been developed, which can be mainly categorized into two groups: (a) deterministic optimization methods including contrast source inversion [10-11] and distorted Born/Rytov iterative methods [12-13], and (b) stochastic methods [14-16] including genetic algorithms and particle swarm optimization algorithms, and so on. Recently, with the emergence of compressive sensing theory, some sparseness-aware inverse scattering algorithms were proposed to mitigate the ill-posedness of underlying inverse problem [17, 43]. Although these methods can produce acceptable results for scenes with moderate size and contrast, it remains an outstanding challenge to deploy them in large and realistic scenes due to the very expensive computational costs. Till now, it has been a consensus that the nonlinear EM inverse scattering technique is mostly limited to the low frequency regime, and has been impeded from many important high-frequency applications, especially in treating the high-contrast objects with strong multi-scattering effects.

In the past few years, deep learning has consolidated as one of the most powerful approaches in several areas of regression and classification problems, due to easy availability of the vast amounts of data and ever-increasing computational power [18-19]. Deep neural network (DNN) approaches have attracted increased attention in image processing and computer vision, such as semantic segmentation [20], depth estimation [21], image deblurring [22], and image super-resolution [23-24]. The DNN approach was also demonstrated to be advantageous over traditional machine learning approaches in the automated analysis of the high-content microscopy data [25]. Deep leaning approach was shown to aid the design and realization of advanced functional materials [26] and high-accuracy reconstruction from compressed measurements [27-28] as well. Most recently, DNN algorithms have been applied in biomedical imaging (e.g., magnetic resonance imaging and X-ray computed tomography) [29-30] and computational optical imaging [7,31-32]. It has been empirically found that the NN-based [33,34] and DNN-based strategies can outperform conventional image reconstruction techniques in terms of improved image quality and reduced computational costs [29-34].

In this work, we established a fundamental connection between a DNN architecture and iterative methods utilized for the nonlinear EM inverse scattering problems. Inspired by this connection, we then develop a novel DNN architecture tailored for the nonlinear EM inverse scattering, which we term 'DeepNIS'. DeepNIS consists of a cascade of multi-layer complex-valued residual CNN modules, which serve to approximately characterize the multi-scattering physical mechanism. The complex-valued residual CNN module is a straightforward extension of the conventional real-valued CNN [23], which is an end-to-end map from an input rough image to the refined solution of a nonlinear inverse scattering problem. The input data of the first module of DeepNIS comes from the back-propagation (BP) image. For the remaining modules of DeepNIS, the input of CNN module is the output of last module. This makes DeepNIS a non-iterative solver, which greatly reduces the computational costs compared to iterative techniques.

The performance of DeepNIS is validated by several proof-of-concept numerical and experimental demonstrations. We train and test the DeepNIS using MNIST dataset (see Appendix b). We also examine its generalization capabilities using the Fresnel experimental data set [41]. We demonstrate that DeepNIS can significantly outperform conventional nonlinear inverse scattering techniques in terms of both image quality and computational time. Specifically, it is shown that DeepNIS is a promising tool for efficiently tackling nonlinear inverse scattering problems including large scenes and high-contrast objects, which is impractical to be solved by using conventional methods.

## II. PROBLEM STATEMENT

We begin our discussion by unveiling the connection between the DNN architecture of interest and iterative methods for nonlinear EM inverse scattering. Since the iterative solution of a nonlinear inverse EM scattering requires convolutions and should account for nonlinearities, this suggests that DNN may offer an efficient alternative solution.

*II.A. Connection between DNN and nonlinear EM inverse scattering*

With reference to the measurement configuration in Fig. 1, we illustrate our strategy in the context of a 2D multiple-input multiple-output (MIMO) measurement configuration. The investigation domain denoted by $D_{inv}$ (inaccessible region), into which the object of interest falls, is successively illuminated by TM-polarized incident waves $E_{inc}^{(n)}$, $n = 1, 2, \ldots, N$ (with $n$ being the index of the $n$th illumination, $N$ is the total number of transmitters). The transmitters and receivers are both located in the observation domain denoted by $\Gamma$ and exterior to $D_{inv}$. For each illumination, the $M$ receivers uniformly distributed over $\Gamma$ are used to collect the electric fields scattered from the probed scene. The time dependence factor $\exp(-i\omega t)$ with angular frequency $\omega$ is used and suppressed throughout the paper. For the $n$th illumination and the $m$th ($m=1, 2, \ldots, M$) receiver, the scattered electrical field $E_{sca}^{(n)}$ at the location of $r_m$ is governed by a pair of coupled equations [10-16]:

$$E_{sca}^{(n)}(\mathbf{r}_m) = k_0^2 \int_{D_{inv}} G(\mathbf{r}_m,\mathbf{r}')\chi(\mathbf{r}')E^{(n)}(\mathbf{r}')d\mathbf{r}' \quad (1)$$

and 
$$E^{(n)}(\mathbf{r}) - E_{inc}^{(n)}(\mathbf{r}) = k_0^2 \int_{D_{inv}} G(\mathbf{r},\mathbf{r}')\chi(\mathbf{r}')E^{(n)}(\mathbf{r}')d\mathbf{r}', \quad (2)$$

$$\mathbf{r}, \mathbf{r}' \in D_{inv}$$

where $\mathbf{r} = (x, y)$ and $\mathbf{r}' = (x', y')$ denote the field and source points, respectively, $E^{(n)}$ represents the total electric field resultant from the interaction of probed scene with incident field $E_{inc}^{(n)}$. $G(\mathbf{r},\mathbf{r}') = \frac{i}{4}H_0^{(1)}(k_0|\mathbf{r}-\mathbf{r}'|)$ denotes the 2D Green's function in free space, where $H_0^{(1)}$ is the first-kind zeroth-order Hankel function. Additionally, the contrast function is defined as $\chi = k^2/k_0^2 - 1$, where $k$ and $k_0$ are the wavenumbers of the probed sample and background medium, respectively.

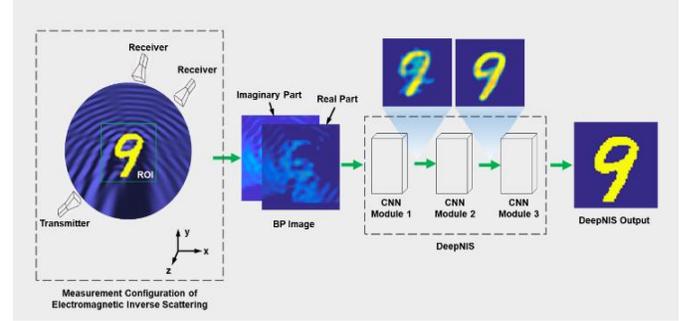

**Fig.1**. Basic configuration of a EM nonlinear inverse scattering problem and the developed DeepNIS solver. Here, two receivers are employed to collect the EM scattering data arising from one transmitter. DeepNIS consists of a cascade of three CNN modules, where the complex-valued input, shown by its real and imaginary parts in this figure, comes from the back-propagation algorithm, and the output is the super-resolution image of EM inverse scattering. Here, the lossless dielectric object is in the shape of a digit "9" and has relative permittivity $\varepsilon_r = 3$.

For computational imaging, the investigation domain $D_{inv}$ is uniformly divided into pixels such that the total electric fields, the contrast currents, and the contrast functions are assumed uniform in each pixel. As a consequence, the nonlinear EM inverse scattering amounts to solving the following coupled equations [10-16]:

$$\mathbf{E}_{sca}^{(n)} = \mathbf{G}_d \mathbf{E}^{(n)} \boldsymbol{\chi} \quad (3)$$

and 
$$\mathbf{E}^{(n)} - \mathbf{E}_{inc}^{(n)} = \mathbf{G}_s \mathbf{E}^{(n)} \boldsymbol{\chi} \quad (4)$$

To solve Eqs. (3) and (4), iterative strategies can be applied. Put formally, the contrast function at the ($k$+1)-iteration step can be obtained by solving the following equation [3]

$$\boldsymbol{\chi}_{(k+1)} = \arg\min_{\chi}\left[\sum_n \left\|\delta \mathbf{E}_{sca}^{(n)} - \mathbf{J}_{(k)}^{(n)}\delta\boldsymbol{\chi}\right\|_2^2 + \Re(\boldsymbol{\chi})\right] \quad (5)$$

where $\delta\mathbf{E}_{sca}^{(n)} \equiv \mathbf{E}_{sca}^{(n)} - \mathbf{E}_{sca}^{(n)}(\boldsymbol{\chi}_{(k)})$ and $\delta\boldsymbol{\chi} \equiv \boldsymbol{\chi} - \boldsymbol{\chi}_{(k)}$. Here, $\boldsymbol{\chi}_{(k)}$ denotes the contrast function evaluated at the $k$-iteration step. Correspondingly, $\mathbf{E}_{sca}^{(n)}(\boldsymbol{\chi}_{(k)})$ denotes the scattered electrical field calculated from the estimation $\boldsymbol{\chi}_{(k)}$ for the $n$th illumination, while $\mathbf{J}_{(k)}^{(n)}$ corresponds to the Jacobian matrix of $\mathbf{E}_{sca}^{(n)}$ with respect to $\boldsymbol{\chi}_{(k)}$. Further, we have introduced the regularization term $\Re(\chi)$ in Eq. (5) to incorporate the a *prior* on the contrast function in order to address the inherent ill-posedness of electromagnetic inverse scattering.

In the area of image processing, it has become a consensus that most of natural images have some structure. This underlying structure allows for a sparse representation in some transformed domain, which also assist on regularization. A properly chosen sparse representation facilitates better image reconstruction [44-46]. Actually, several sparsity-aware electromagnetic inverse scattering methods have been developed recently [17, 44-46]. Here, for simplicity, we consider $\Re(\chi) = \|\mathbf{D}\chi\|_1$, where $\mathbf{D}$ denotes a specified sparse transformation, like wavelet, etc. As a consequence, after employing so-called proximal approximation technique, we can arrive at the solution to Eq. (5) as follows [45]

$$\chi_{(k+1)} = \mathbf{D}^H \mathcal{S}\left\{\mathbf{D}\chi_{(k)} + \mathbf{D}\left[\sum_n \left(\mathbf{J}_{(k)}^{(n)}\right)^H \mathbf{J}_{(k)}^{(n)}\right]^\dagger \sum_n \left(\mathbf{J}_{(k)}^{(n)}\right)^H \delta\mathbf{E}_{sca}^{(n)}\right\}$$

(6)

Herein, $\mathcal{S}\{\cdot\}$ denotes the element-wise soft-threshold function, and the subscript *H* denotes the conjugate transpose.

In order to make the connection between DNN and the iterative solution to a nonlinear electromagnetic inverse scattering, we rearrange Eq. (6) into the following form

$$\begin{aligned}
\mathbf{D}\chi_{(k+1)} &= \mathcal{S}\left\{\mathbf{D}\chi_{(k)} + \mathbf{D}\left[\sum_n \left(\mathbf{J}_{(k)}^{(n)}\right)^H \mathbf{J}_{(k)}^{(n)}\right]^\dagger \sum_n \mathbf{A}_{(k)}^{(n)} \delta\mathbf{E}_{sca}^{(n)}\right\} \\
&= \mathcal{S}\left\{\mathbf{D}\chi_{(k)} + \mathbf{D}\left[\sum_n \left(\mathbf{J}_{(k)}^{(n)}\right)^H \mathbf{J}_{(k)}^{(n)}\right]^\dagger \sum_n \mathbf{A}_{(k)}^{(n)} \left(\mathbf{E}_{sca}^{(n)} - \mathbf{G}_d \mathbf{E}_{(k)}^{(n)} \chi_{(k)}\right)\right\} \\
&= \mathcal{S}\left\{\mathbf{P}_{(k)} \chi_{(k)} + \mathbf{b}_{(k)}\right\}
\end{aligned}$$

(7)

where $\mathbf{P}_{(k)} \equiv \mathbf{D}\mathbf{P}_{(k)} - \mathbf{D}\mathbf{P}_{(k)}\left[\sum_n \left(\mathbf{J}_{(k)}^{(n)}\right)^H \mathbf{J}_{(k)}^{(n)}\right]^\dagger \sum_n \mathbf{A}_{(k)}^{(n)} \mathbf{G}_d \mathbf{E}_{(k)}^{(n)}$

and $\mathbf{b}_{(k)} \equiv \mathbf{D}\left[\sum_n \left(\mathbf{J}_{(k)}^{(n)}\right)^H \mathbf{J}_{(k)}^{(n)}\right]^\dagger \sum_n \mathbf{A}_{(k)}^{(n)} \mathbf{E}_{sca}^{(n)}$.

Note that $\mathbf{E}_{(k)}^{(n)}$ defined in the second line of Eq. (7) represents the total electrical field inside the domain of interest. The recursive solution (7) resembles that of full-connected deep neural network. In the terminology of deep learning, $\mathbf{P}_{(k)}$ and $\mathbf{b}_{(k)}$ can be understood as the weighting matrix and the bias, respectively. Likewise, the iterative index *k* corresponds to the layer index of deep neural network, while the soft-threshold function $\mathcal{S}\{\cdot\}$ corresponds to the nonlinear activation function in deep learning.

Invoked by deep learning [27, 39, 40], when a set of samples are available at hand, it is appealing to train both $\mathbf{P}_{(k)}$ and $\mathbf{b}_{(k)}$ for each layer. Comparing this approach to conventional iterative inverse scattering methods, the expectation is that the learned method would be more efficient as it optimizes the weighting matrices and biases, and targets the reconstruction error with respect to the ground-truth images. In summary, above observations suggest that deep neural networks are naturally well-suited for nonlinear EM inverse scattering problems. It is worth remarking that the resulting DNN architecture differs from the conventional DNNs in the sense that it is complex-valued rather than real-valued.

*II.B. Deep DNN for nonlinear EM inverse scattering*

After demonstrating the natural connection between the deep DNN architecture and nonlinear EM inverse scattering, we now develop a complex-valued deep DNN (i.e., DeepNIS) to solve the nonlinear EM inverse scattering problem. For the sake of DNN computational complexity, DeepNIS can be designed as a cascade of CNN modules, as shown in Fig. 1, where the input data of DeepNIS comes from the back-propagation (BP) image. For the remaining modules of DeepNIS, the output of last CNN module is the input of the next module. Each CNN module consists of several up-sampling convolution layers and each up-sampling convolution layer consists of three steps: in the first step, the input is convolved with a set of learned fitters, resulting in a set of feature (or kernel) maps; in the second step, these maps undergo a point-wise nonlinear function, resulting in a sparse outcome; an optional third down-sampling step (termed as pooling) is applied on the result to reduce its dimensions, thus forming the multi-layer structure. More details can be found in **Appendix A**.

III. NUMERICAL AND EXPERIMENTAL RESULTS

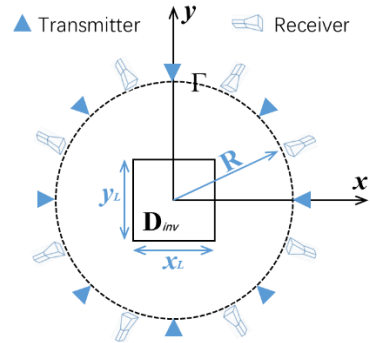

**Fig.2.** Measurement configuration for the electromagnetic inverse problem scenario.

In the following, we numerically and experimentally evaluate the performance of DeepNIS in solving nonlinear EM inverse scattering problems. For comparison, we also report corresponding results by using the contrast source inversion (CSI) method, which has been popularly used in nonlinear inverse scattering. The discrete dipole method is used to generate the simulation data.

*III.A Training and testing over MNIST dataset*

We train and test the DeepNIS using MNIST dataset, which is a database of ten handwritten digits from 0 to 9 and has been widely used in machine learning (see **Appendix B**).

With reference to Fig. 2, the region of interest $D_{inv}$ is a square with size of $5.6 \times 5.6 \lambda_0^2$ ($\lambda_0$=7.5 cm is the working wavelength in vacuum and $x_L=y_L=5.6\lambda_0$), which is uniformly divided into 110×110 sub-squares for the simulations. Moreover, 36 linearly polarized transmitters, which are located uniformly over the circle denoted by $\Gamma$ with radius R=10 $\lambda_0$, successively illuminate the investigation domain. Meanwhile, 36 co-polarized receivers, are used to simultaneously collect the electrical field scattered from the probed scene. In the full-wave EM simulations [38], the digit-like objects are set to be lossless dielectrics with a relative permittivity of $\varepsilon_r = 3$. In addition, 30 dB noise has been added for all simulations throughout this article to avoid the so-called "inverse crime". Note that we train the DeepNIS only in the noiseless case. A total of $10^4$ images are randomly chosen from the MNIST dataset as samples. And the multi-input and multi-output EM responses are obtained by running a full-wave solver to the Maxwell's equations. As a result, $10^4$ back propagation (BP) images can be generated, which are used as inputs to DeepNIS, while the original $10^4$ images are considered as the desirable outputs in DeepNIS. Meanwhile, $10^4$ image pairs are randomly divided into three sets: 7000 image pairs for training, 1000 image pairs for validation, and other 2000 image pairs for blind testing.

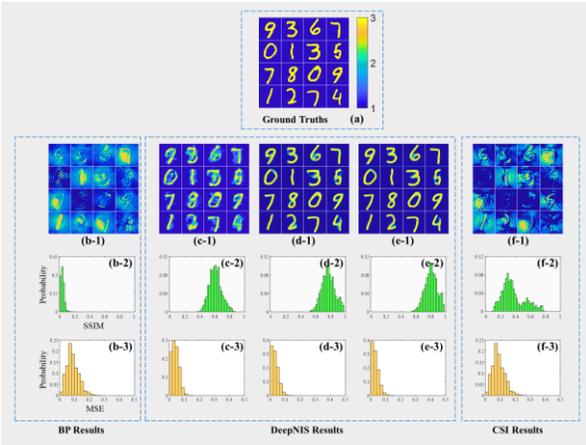

**Fig.3**. Reconstructions of digit-like objects with relative permittivity $\varepsilon_r = 3$ by different EM inverse scattering methods. (a) Sixteen ground truths. (b-1) BP results, which are used as the input of DeepNIS. (c-1, d-1, e-1) DeepNIS results with different numbers of CNN modules, viz., 1, 2, 3, respectively. (f-1) CSI results. (b-2) to (f-3) Statistical histograms of the image quality in terms of SSIM and MSE shown in the third and fourth line in Fig. 3, respectively. Here, 2000 test samples are used in the statistical analysis. For visualization purpose, the BP reconstructions are normalized by their own maximum values, since their values are much less than 1.

The networks are trained using ADAM optimization method [42], with mini-batches size of 32, and epoch setting as 101. The learning rates are set to $10^{-4}$ and $10^{-5}$ for the first two layers and the last layer in each network and divided by 2 when the error plateaus. The complex-valued weights and biases are initialized by random weights with Gaussian distribution of zero mean and standard deviation of $10^{-3}$. With a Euclidean cost, these networks are trained independently, but finally, tuned in an end-to-end manner. All computations are performed in a small-scale server with the configuration of 128GB access memory, Intel Xeon E5-1620v2 central processing unit, NVIDIA GeForce GTX 1080Ti. The deep learning networks are both designed with Tensorflow library [43] and CSI algorithms are carried out by Matlab 2017. And the networks training takes about 7 hours.

Figure 3(a) represents the ground truths for the simulated ten handwritten digits in the nonlinear inverse problem. Figs. 3(b-1) and 3(f-1) report the images obtained by using the BP algorithm and the CSI method, respectively, which clearly illustrates that both the BP and the CSI fail to produce the satisfactory reconstructions in this case. Figures 3(c-1), (d-1) and (e-1) provide the corresponding results calculated by the DeepNIS with 1, 2, and 3 CNN modules, respectively.

In order to investigate the effects of the number of CNN-modules on the image quality. We adopt the so-called *Structure Similarity Measure* (SSIM) and *Mean-Square Error* (MSE) as qualitative measure metrics to evaluate the image quality. Figs. 3(b-2), (c-2), (d-2), (e-2) and (f-2) report the statistical histograms of the image quality in terms of SSIM, corresponding to Figs. 2(b-1), (c-1), (d-1), (e-1) and (f-1), respectively, over 2000 test images, where the *y*-axis is normalized to the total 2000 test images. It can be clearly seen that the DeepNIS results obtained with 2 or 3 CNN modules could almost perfectly match the ground truth results. It is worth mentioning that it only takes a well-trained DeepNIS less than one second to construct an image in this case, whereas it takes BP and CSI algorithm about 8 seconds and about 10 minutes, respectively. A similar conclusion can be draw from the results of MSE index. Based on the above results, it can be concluded that the DeepNIS clearly outperforms the CSI method in terms of both image quality and computation time in this high-contrast case. In addition, it is expected that the use of additional CNN modules will enable incorporation of more multiple scattering effects into account, leading to an improved image quality.

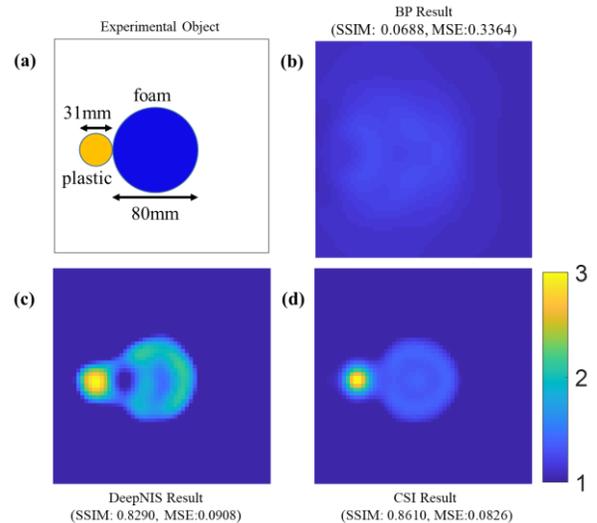

**Fig.4.** Experimental reconstructions by different EM inverse scattering methods. (a) The probed object consists of a composition of cylindrical foam

(blue) and plastic (yellow) objects. (b, c, d) Reconstruction results using BP, DeepNIS, and CSI methods. The corresponding SSIMs (MSE) of the reconstructed images are equal to 0.0668(0.3364), 0.8290(0.0908), and 0.8637(0.0826), respectively.

*III.B Testing over experimental data with trained networks*

To investigate the generalizability of DeepNIS, we consider the *FoamDielExt* experimental data provided by the Institute Fresnel, Marseille, France [41] with the CNNs trained through the MNIST dataset. The configuration of the experimental measurement setup has been carefully described in [41]. For numerical simulations, the investigation domain is uniformly divided into 56×56 sub-squares. Figure 4(a) shows the *FoamDielExt* object (ground truth), where the yellow object is a dielectric (plastic) with a relative permittivity of 3 ±0.3, and the blue object is a dielectric (foam) with a relative permittivity of 1.45 ± 0.15. The working frequency is 4 GHz. The results produced by the BP algorithm, the CSI method, and DeepNIS are shown in Figs. 4(b), (c), and (d), respectively. Although the ground truth in this case is remarkably different from the training samples of the MNIST dataset, the result obtained by DeepNIS is satisfied and comparable to that of CSI. It should be pointed out that DeepNIS is several orders of magnitude faster than the CSI method. Specifically, it takes DeepNIS around 1 second to produce this results, but it costs CSI several minutes and 70 iterations.

Note that the dielectric contrast of the object in this experimental test is low, which corresponds to the range of validity of the CSI method. However, as shown in Fig. 3, if the test object has a high contrast, the CSI method fails to adequately reconstruct the image due to stronger multiple scattering effects. In contrast, DeepNIS is expected to perform well in that regime as well.

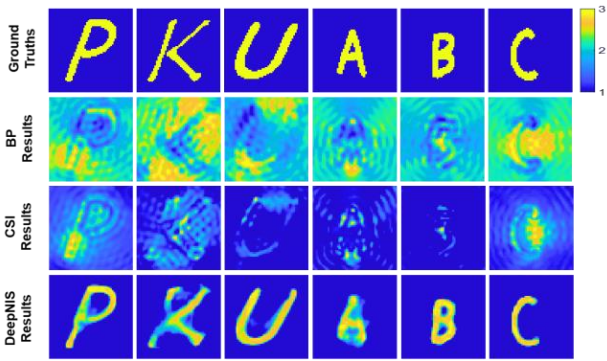

**Fig.5**. Reconstruction results of letter-shaped objects by the BP algorithm, the CSI method, and DeepNIS in the second, third, and fourth rows, respectively. The ground truth is shown in the first row.

*III.C Testing over letter targets with trained networks*

In order to validate above points, we conduct another set of simulations, in which DeepNIS is still trained over the MNIST dataset. The test objects are composed of dielectric shapes in the form of English letters, whose relatively permittivity is 3. Other parameters are all as same as training dataset.

Figure 5 shows the reconstructed results based on different inverse scattering methods, in which the ground truths are given in the first row, and the imaging results by the BP algorithm, CSI, and DeepNIS are presented in the second, third, and fourth rows, respectively. To compare the imaging quality in the reconstruction of English-letter objects using the BP algorithm, the CSI method, and DeepNIS, the corresponding SSIM and MSE results from different methods are respectively reported in Table 1 and Table 2. Meanwhile, the reconstructed procedure with trained network in Example 1 just take less than 1 second, while the CSI algorithm needs 50 iterations and take about 10 minutes for reconstruction. The BP algorithm also has relatively lower computational complexity and takes about 8 seconds. Since the probed objects have large contrasts, the CSI method fails to provide acceptable images. The reconstruction results clearly demonstrate that DeepNIS is markedly superior to both BP algorithm and CSI method in both imaging quality and imaging time.

Table Ⅰ
SSIM results for the reconstructions in Fig.5

| Target | P | K | U | A | B | C |
|---|---|---|---|---|---|---|
| BP | 0.0172 | 0.0078 | 0.0111 | 0.0871 | 0.0518 | 0.0987 |
| CSI | 0.4649 | 0.2308 | 0.3982 | 0.2347 | 0.2321 | 0.2367 |
| DeepNIS | 0.8912 | 0.6365 | 0.9192 | 0.8775 | 0.9473 | 0.9395 |

Table Ⅱ
MSE results for the reconstructions in Fig.5

| Target | P | K | U | A | B | C |
|---|---|---|---|---|---|---|
| BP | 0.7571 | 0.8319 | 0.8467 | 0.8491 | 0.8813 | 0.8401 |
| CSI | 0.5518 | 0.6781 | 0.4829 | 0.7104 | 0.8492 | 0.5199 |
| DeepNIS | 0.0542 | 0.0689 | 0.0281 | 0.0934 | 0.0118 | 0.0120 |

From above discussions, we can arrive at an important conclusion: despite the fact that the network was trained exclusively on images from the MINIST dataset, satisfactory reconstruction results can still be obtained from very different objects by using the trained DeepNIS. This suggests that the DeepNIS has learned a model of the underlying physics of the imaging system or at least a generalizable mapping between the input BP results and the output inverse scattering solutions when training and testing dataset in similar electromagnetic inverse scattering scenario. We clearly observe that the DeepNIS images have a considerably higher SSIM than the BP and CSI images. In other words, these results suggest the DeepNIS is not merely matching patterns but has actually has a learning capability to represent the underlying nonlinear inverse electromagnetic scattering problem.

## IV. CONCLUSIONS

In conclusion, we have built up a connection between CNN and unfolded iterative solution to nonlinear EM inverse scattering, and then established a complex-valued DNN, termed as DeepNIS, for the non-iterative solution of nonlinear EM inverse scattering problems. A central issue to the DeepNIS-based solution is the convolution operation, which can be implemented in parallel. The non-iterative and parallelizable natures of DeepNIS make it very suitable for dealing with large-scale inverse scattering problems. We showed that DeepNIS has clear advantages over conventional inverse scattering methods in terms of image quality and computational time. Our experimental results suggest that the DeepNIS can "learn" the governing equations of the electromagnetic inverse scattering system, when training and testing dataset in similar electromagnetic inverse scattering scenario. It is plausible that more advanced CNN architectures may yield even better results, which would be explored in our further study. DeepNIS could improve upon conventional inverse scattering strategies, and be used for treating the nonlinear EM inverse scattering with large scale and high contrast objects.

## APPENDIX A. THE COMPLEX-VALUED CNN MODULE OF DEEPNIS

A complex-valued CNN module of DeepNIS contains three layers (Fig. A1): an up-sampling convolution layer followed by a nonlinear activation function, a max-pooling layer, and an up-sampling layer. The up-sample convolutional layer is expressed as the operation:

$$F_1(Y) = \text{ReLU}(W_1 * Y + B_1) \quad \text{(A1)}$$

where $W_1$ and $B_1$ represent the complex-valued filters and biases, respectively. $*$ denotes the convolution operation and ReLU denotes the rectified linear unit activation function. $Y$ means the images of input. Here, $W_1$ corresponds to $n_1$ filters of the support $f_1 \times f_1$ in which $f_1$ is the spatial size of a filter. The last layer is the convolution layer for reconstruction:

$$F_3(Y) = \text{ReLU}(W_3 * Y + B_3) \quad \text{(A2)}$$

where $W_3$ and $B_3$ represent the complex-valued filters with a size of $f_3 \times f_3$ and biases, respectively.

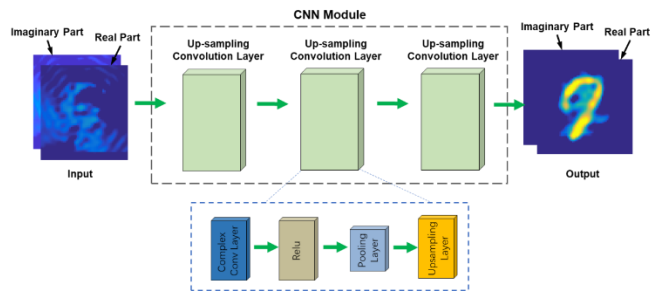

**Fig.A1.** Schematic illustration of the first CNN module of DeepNIS.

Given an object, its relative permittivity and conductivity are assumed to be non-negative. Considering this fact, the activation function ReLU is used throughout this article. Note that ReLU is separately operated on the real and imaginary part of underlying complex-valued input. For each module, three layers are enough to achieve the desired image quality in all cases we considered. If needed, more convolutional layers can be added to enrich the nonlinearity of the undergoing system; however, this increases the complexity of the model, and thus demands extra training time and increases the risk of overfitting.

## APPENDIX B. MNIST DATASET

In our numerical study, the probed objects are modeled by exploring MNIST, a dataset of handwriting digits widely used in the area of machine learning [47]. For the electromagnetic simulations, the objects are set to be lossless dielectrics with relative permittivity of 3. Some MNIST samples are shown in **Fig. A2**.

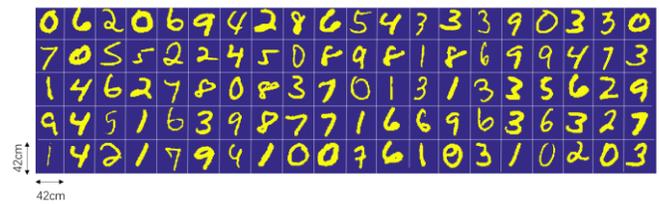

**Fig. A2.** Some MNIST samples used in Figs.3-5.

ACKNOWLEDGMENT

This research is funded in part by National Natural Science Foundation of China (NSF) grant 61471006), the 111 Project grant 111-2-05, and the AFOSR grant FA9550-16-1-0386.

## REFERENCES

[1] W. Kofman, A. Herique, Y. Barbin, Jean-Pierre Barriot, V. Ciarletti, S. Clifford, P. Edenhofer, C. Elachi, C. Eyraud, Jean-Pierre Goutail, E. Heggy, L. Jorda, J. Lasue, Anny-Chantal Levasseur-Regourd, E. Nielsen, P. Pasquero, F. Preusker, P. Puget, D. Plettemeier, Y. Rogez, H. Sierks, C. Statz, H. Svedhem, I. Williams, S. Zine, and J. Van Zyl, Properties of the 67P/Churyumov-Gerasimenko interior revealed by CONSERT, radar. *Science*, vol. 349, no. 6247, pp. aab0639, 2015.

[2] A. Redo-Sanchez, B. Heshmat, A. Aghasi, S. Naqvi, M. Zhang, J. Romberg, and R. Raskar. Terahertz time-gated spectral imaging for content extraction through layered structures. *Nat. Commun.*, 7, 12665, 2016.

[3] M. Pastorino. Microwave imaging, John Wiley & Sons, Press, 2010.

[4] D. Colton and R. Kress. Inverse Acoustic and Electromagnetic scattering theory. 93, *Springer Science & Business Media*, 2012.

[5] G. Maire, F. Drsek, J. Girard, H. Giovannini, A. Talneau, D. Konan, K. Belkebir, P. C. Chaumet, and A. Sentenac. Experimental demonstration of quantitative imaging beyond Abbe's limit with optical diffraction tomography, *Phys. Rev. Lett.* 102, 213905, 2009.

[6] P. M. Meaney, M. W. Fanning, D. Li, S. P. Poplack, and K. D. Paulsen, A clinical prototype for active microwave imaging of the breast. *IEEE Trans. Microwave Theory Technique*, 48(11), 1841-1853, 2000

[7] L. Waller, and L. Tian. Computational imaging: machine learning for 3D microscopy. *Nature*, 523, 416-417, 2015.

[8] O. Haeberlé, K. Belkebir, H. Giovaninni, and A. Sentenac. Tomographic diffractive microscopy: basics, techniques and perspectives. *J. Mod. Opt.* 57, 686–699, 2010.

[9] A. C. Kak and M. Slaney. Principles of computerized tomographic imaging, *SIAM Press*, 2001

[10] P. M. van den Berg, and A. Abubakar. Contrast source inversion method: state of art, *Progress in Electromagnetic Research*. 34, 189-218, 2001.

[11] L. Li, H. Zheng, and F. Li. Two-dimensional contrast source inversion method with phaseless data: TM case. *IEEE Trans. Geoscience and Remote Sensing*, 47(6), 1719-1735, 2009.

[12] W. C. Chew and Y. M. Wang. Reconstruction of two-dimensional permittivity distribution using the distorted Born iterative method. *IEEE Trans. Medical Imaging*, 9(2), 218-225, 1990.

[13] L. Li, L. G. Wang, J. Ding, P. K. Liu, M. Y. Xia, and T. J. Cui, A probabilistic model for the nonlinear electromagnetic inverse scattering: TM case. *IEEE Trans. Antenna and Propagation*, 65(11):5984-5991, 2017.

[14] P. Rocca, M. Benedetti, M. Donelli, D. Franceschini, and A. Massa. Evolutionary optimization as applied to inverse scattering problems. *Inverse Problems,* 25(12), 123003, 2009

[15] P. Rocca, G. Oliveri, and A. Massa. Differential evolution as applied to electromagnetics. *IEEE Antennas Propagation Magazine*, 53(1), 38–49, 2011.

[16] M. Pastorino. Stochastic optimization methods applied to microwave imaging: A review. *IEEE Trans. Antennas Propagation*, 55(3), 538–548, 2007.

[17] N. Anselmi, M. Salucci, G. Oliveri, and A. Massa, Wavelet-based compressive imaging of sparse targets, *IEEE Trans. Antenna and Propagation*, vol. 63, no. 11, pp. 4889-4899, 2015.

[18] Y. LeCun, Y. Bengio, and G. Hinton. Deep learning. *Nature*. 521, 28, 436-444, 2015.

[19] I. Goodfellow, Y. Bengio, and A. Couriville. Deep learning. Cambridge: MIT Press, 2016.

[20] J. Long, E. Shelhamer, and T. Darrell. Fully convolutional networks for semantic segmentation. *IEEE Conf. Comp. Vision and Pattern Recog.*, 1, 3, 2015.

[21] D. Eigen, C. Puhrsch, and R. Fergus. Depth map prediction from a single image using a multi-scale deep network. *Adv. Neural Inf. Proc. Sys.* 2366-2374, 1, 3, 2014.

[22] J. Sun. W. Cao, Z. Xu, and J. Ponce. Learning a convolutional neural network for non-uniform motion blur removal. *Proceedings of IEEE Conference on Computer Vision and Pattern Recognition*, 1, 2, 2015.

[23] C. Dong, C. Loy, K. He, and X. Tang. Image super-resolution using deep convolutional networks. 38, 2, 296-307, 2016.

[24] D. Liu, Z. Wang, B. Wen, J. Yang, W. Han, and T. S. Huang. Robust single image super-resolution via deep network with sparse prior. IEEE Trans. Image Processing, 25(7), 3194-3207, 2016.

[25] O. Z. Kraus, B. T. Grys, J. Ba, Y. Chong, B. J. Frey, C. Boone, and B. J. Andrews. Automated analysis of high-content microscopy data with deep learning. *Molecular Systems Biology*, 32, 924, 2017.

[26] S. V. Kalinin, B. G. Sumpter, and R. K. Archibald. Big-deep-smart data imaging for guiding materials design. *Nature Materials*, 14, 4395, 973-980, 2015.

[27] A. Mousavi, and R. Baraniuk. Learning to invert: signal recovery via deep convolutional networks. *ICASP*, 2017.

[28] K. Kulkarni, S. Lohit, P. Turaga, et al., ReconNet: Non-iterative reconstruction of images from compressively sensed measurements. IEEE conference on computer vision and pattern recognition (CVPR), 2016

[29] Y. Han, J. Yoo, and J. C. Ye. Deep residual learning for compressed sensing CT reconstruction via persistent homology analysis. [Online]. Available: https://arxiv.org/abs/1611.06391, 2016.

[30] Kyong Hwan Jin, Michael T. McCann, E. Froustey, and M. Unser. Deep convolutional neural network for inverse problems in imaging. *IEEE Trans. Signal Processing,* 26(8):4509-4522. 2017.

[31] A. Sinha, J. Lee, S. Li, and G. Barbastathis. Lensless computational imaging through deep learning. *Optica*, 4, 9, 1117-1125, 2017.

[32] U. S. Kamilov, I. N. Papadopoulos, M. H. Shoreh, A. Goy, C. Vonesch, M. Unser, and D. Pasaltis. Learning approach to optical tomography. *Optica*, 2(6), 517-522, 2015.

[33] Q. Marashdeh, W. Warsito, L.-S. Fan, and F. L. Teixeira, Nonlinear forward problem solution for electrical capacitance tomography using feed forward neural network, *IEEE Sensors J.*, 6(2), 441-449, 2006.

[34] Q. Marashdeh, W. Warsito, L.-S. Fan, and F. L. Teixeira, A nonlinear image reconstruction technique for ECT using combined neural network approach, Meas. Sci. Technol., 17(8), 2097-2103, 2006

[35] L. Di Donato, M. T. Bevacqua, L. Crocco and T. Isernia, Inverse Scattering Via Virtual Experiments and Contrast Source Regularization, *IEEE Transactions on Antennas and Propagation*, 63, 4, 1669-1677, 2015

[36] L. Di Donato, R. Palmeri, G. Sorbello, T. Isernia and L. Crocco, A New Linear Distorted-Wave Inversion Method for Microwave Imaging via Virtual Experiments, *IEEE Transactions on Microwave Theory and Techniques*, 64, 8, 2478-2488, 2016.

[37] R. Palmeri, M. T. Bevacqua, L. Crocco, T. Isernia and L. Di Donato, Microwave Imaging via Distorted Iterated Virtual Experiments, *IEEE Transactions on Antennas and Propagation*, 65, 2, 829-838, 2017

[38] M. F. Catedra, R. P. Torres, J. Basterrechea, and E. Gago. The CG-FFT method: application of signal processing techniques to electromagnetics. Boston, MA: Artech House, 1995.

[39] K. Gregor, and Y. LeCun, Learning fast approximations of sparse coding, In *ICML*, 399–406, 2010

[40] Z. Wang, D. Liu, J. Yang, W. Han and T. Huang, Deep Networks for Image Super-Resolution with Sparse Prior, *IEEE International Conference on Computer Vision (ICCV)*, Santiago, Chile, 370-378, 2015

[41] J.-M. Geffrin, P. Sabouroux, and C. Eyraoud. Free space experimental scattering database continuation: Experimental set-up and measurement precision. *Inv. Probl.*, 21(6): S117–S130, 2005.

[42] D. P. Kingma and J. Ba, `Adam: A method for stochastic optimization,' arXiv:1412:6980, 2017.

[43] M. Abadi, A. Agarwal, P. Barham, E. Brevdo, Z. Chen, C. Citro, G. S. Corrado, A. Davis, J. Dean, M. Devin, S. Ghemawat, I. Goodfellow, A. Harp, G. Irving, M. Isard, R. Jozefowicz, Y. Jia, L. Kaiser, M. Kudlur, J. Levenberg, D. Mané, M. Schuster, R. Monga, S. Moore, D. Murray, C. Olah, J. Shlens, B. Steiner, I. Sutskever, K. Talwar, P. Tucker, V. Vanhoucke, V. Vasudevan, F. Viégas, O. Vinyals, P. Warden, M. Wattenberg, M. Wicke, Y. Yu, and X. Zhen, TensorFlow: Large-scale machine learning on heterogeneous systems, (white paper) 2015. Software available from tensorflow.org.

[44] A. E. Fouda and F. L. Teixeira, Bayesian compressive sensing for ultrawideband inverse scattering in random media, *Inverse Problems*, vol. 30, no. 11, 114017, 2014.

[45] A. E. Fouda and F. L. Teixeira, Ultra-wideband microwave imaging of breast cancer tumors via Bayesian inverse scattering, *Journal of Applied Physics*, vol. 115, no. 6, 064701, 2014.

[46] L. Li, M. Hurtado, F. Xu, B. Zhang, T. Jin, T. Cui, M. Nikolić Stevanović, and A. Nehorai, A survey on the low-dimensional-model-based electromagnetic imaging, *Foundations and Trends in Signal Processing*, 12, 2, 107-199, 2018

[47] Y. LeCun, L. Bottou, Y. Bengio, and P. Haffner, Gradient-based learning applied to document recognition, *Proc. IEEE*, vol. 86, pp. 2278–2324, 1998.